# Gearing of nitrate ions in ammonium nitrate

Na Du,† Xintian Wang,† Yu Ying Zhu, Chanreingam Long, Peng Ren,* Fei Yen*

School of Science, Harbin Institute of Technology, Shenzhen, University Town, Shenzhen, Guangdong 518055, P. R. China.

†These authors contributed equally;
*To whom correspondence can be addressed: fyen@hit.edu.cn or renpeng@hit.edu.cn

**Abstract:** Reorienting polyatomic ions such as $NH_4^+$ and $NO_3^-$ exhibit weak magnetic fields because the ions at the extremities trace out current loops; if the periodic reorientations become long-range ordered (i.e. gearing of neighboring $NO_3^-$), then the magnetic susceptibility should exhibit a unique signature along the different crystallographic axes. For the case of ammonium nitrate $NH_4NO_3$, we report the presence of two successive sharp steps in the molar magnetic susceptibility along the *a*- and *b*-axes upon crossing its order-disorder phase transition (from phase IV to phase II). We suggest the first step pertains to the $NO_3^-$ planes shifting away from facing only along the *b*-axis and onto the *a*-axis by 45°. The second step is attributed to the disordering (ungearing) of the $NH_4^+$ and $NO_3^-$. In contrast, only one step was observed in the magnetic susceptibility along the *c*-axis and its large magnitude suggest the $NO_3^-$ remain weakly correlated even in phase I at 400 K. We also find evidence that the $NH_4^+$ become magnetically ordered (geared) along the *c*-axis only until phase V. The approach employed in this work can be extended to experimentally study the lattice dynamics of other solids possessing planar ions such as amphidynamic crystals.

*The following article has been submitted to The Journal of Chemical Physics.*

INTRODUCTION

Ammonium nitrate (AN) NH$_4$NO$_3$ is rather unique as it possesses at least *five* phases at ambient pressure.[1-15] Figure 1a shows the temperature ranges of phases **I**, **II**, **IV** and **V** and a schematic representation of their equilibrium positions. There is a phase **III** that is observable when there is a presence of small amounts of water[3,12] and a phase **VII** present below 103 K but its crystal structure remains unknown.[2,6] According to general consensus,[4,6,10,14] the higher temperature phases **I** and **II** are treated to be disordered while the lower temperature phases **IV** and **V** as ordered. But there are also reports concluding that **II** is partially ordered[3,16] and that **IV** is partially disordered.[3,17,18] From such, a natural question arises: to what extent are the cations and anions ordered in phases **IV** and **II**?

The degree of ordering of a system can be studied by measurements of the heat capacity and analyzing the excess change in entropy.[19,20] Magnetic susceptibility measurements can also provide the extent of magnetic ordering at different temperatures of systems possessing at least one unpaired electron.[21,22] In the cases when all electrons are paired up, their contribution to the magnetic susceptibility is constant with respect to temperature. The nuclear spins of the atoms in certain systems may also become ordered, but this typically occurs at ultra-low temperatures.[23] Now, there is an intermediate type of magnetic ordering where the carriers are not as mobile as electrons nor as stationary as heavy atoms. In polyatomic ions such as NH$_4^+$, the central N atom is stationary but the protons (H$^+$) are constantly hopping in concert when the entire cation reorients by 90° or 120° (*i.e.* $C_4$ or $C_3$ reorientations illustrated in Fig. 1(b)). Since the four protons in an NH$_4^+$ may be treated to each possess a charge of $+q/4$, whether the reorientation is $C_4$ or $C_3$, a complete current loop is always traced out by the four or three protons hopping in concert to yield an associated magnetic moment $\mu_{NH4}$; from which the direction of $\mu_{NH4}$ points along the direction perpendicular of the loop.[24] The $C_4$ reorientations are more energetically costly because it involves motion of four protons. Below a threshold thermal energy when the $C_4$ reorientations become energetically forbidden, the remaining $C_3$ modes are not enough to accommodate neighboring NH$_4^+$ to possess enough different states. Consequently, the associated $\mu_{NH4}$ of neighboring cations are constantly pointing along the same directions (interacting) so the system is forced to stagger the directions of $\mu_{NH4}$ to reduce the resonant forces. Such ordering, while temporal in nature, also causes the NH$_4^+$ to appear to order spatially.[25,26] During the ordering process $\mu_{NH4}$, the magnetic susceptibility $\chi(T)$ exhibits a step-up discontinuity with decreasing temperature.[25-28]

In the case of the NO$_3^-$ anion, the N and O atoms may be treated to possess charges of near $+q$ and $-2q/3$, respectively; these are rather generic values as different models employ distinct global charges,[29,30] what is important is that the negatively charged oxygen atoms are constantly in motion. The four atoms reside in the same plane so $NO_3^-$ ions easily reorient about the central N atom axis perpendicular to the plane because symmetry can be conserved. Such type of motion generates an associated magnetic moment $\mu_{NO3}$ perpendicular to the planes [Fig. 1(b)]. Even if an $NO_3^-$ ion does not reorient by 120° so that its three $O^{-2/3}$ enclose a loop, weak *ac* magnetic fields are still generated if the anion exhibit in-plane librations.[31] If there is magnetic ordering of such anions, which is rare if contrasted to cations,[32] then $\chi(T)$ should exhibit a step-down (or step-up) discontinuity during the ordering (or disordering) process. At this point, we wish to emphasize that magnetic ordering of $NO_3^-$ *is not based on* conventional electron-electron interactions, instead, it refers to the librations of neighboring $NO_3^-$ becoming synchronized but offset by 180°. In the disordered phase, the $NO_3^-$ librate independently from each other; in the ordered phase, the $NO_3^-$ may also be regarded as being geared, synchronized, correlated or locked in phase (eclipsed) or out of phase (staggered) with respect to each other.

Lastly, when there is a magnetic moment reorientation in conventional systems, pronounced sharp increases and decreases are exhibited along the different crystallographic orientations.[33,34] As AN phase transitions from **IV** to **II**, the directions of the *a*-, *b*- and *c*-axes remain unchanged;[3] mainly the normal vector of the $NO_3^-$ ions shift by 45° along the *ab*-plane so it is interesting to check whether $\chi(T)$ along the *a*- and *b*-axis can also reveal this information.

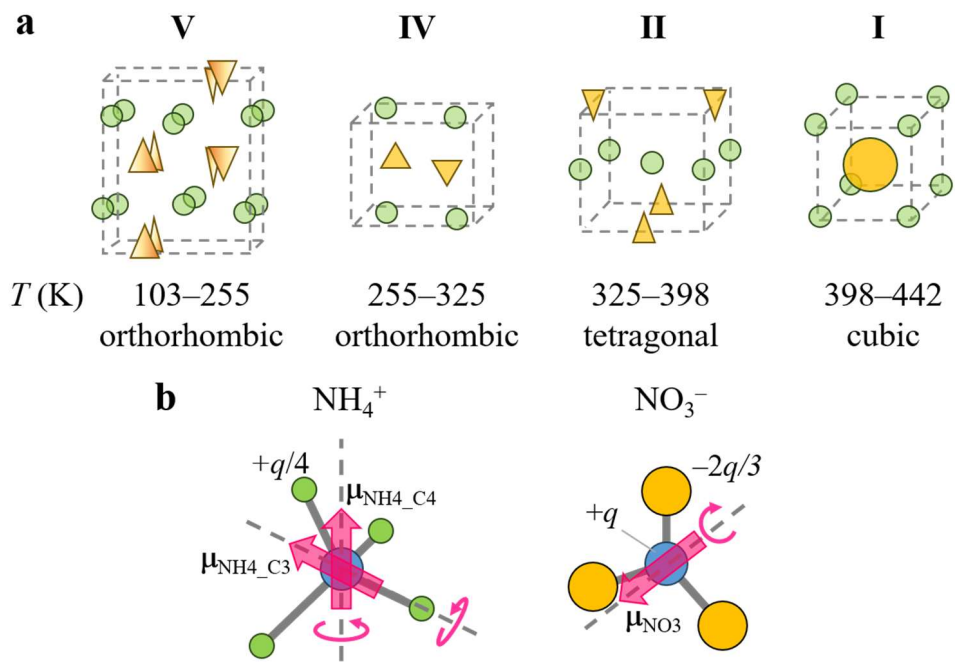

**FIG. 1.** (a) Schematic representation of known structural phases of ammonium nitrate $NH_4NO_3$ and their stability regions. Green and yellow spheres represent $NH_4^+$ and $NO_3^-$. (b) Charge distribution of the $NH_4^+$ and $NO_3^-$ ions. Weak magnetic moments, $\mu_{NH4\_C4}$, $\mu_{NH4\_C3}$, and $\mu_{NO3}$ are generated when an $NH_4^+$ reorients by 90° or 120° and $NO_3^-$ by 60°, respectively.

In this work, we first present the measured magnetic susceptibility of AN single crystals along its three principal crystallographic directions in the range of 300 to 400 K covering the phases **IV**, **II** and **I**. The **IV-II** phase transition is found to be comprised of two sharp steps in succession along the *a* and *b*-axes. The first step is interpreted to pertain to the $NO_3^-$ planes, originally facing the *b*-axis, to begin shifting toward the *a*-axis. The second step is believed to be associated to the disordering of the $NH_4^+$ and $NO_3^-$. The $NH_4^+$ and $NO_3^-$ are verified to be completely disordered in **II** along the *a*- and *b*-axes. However, the $NO_3^-$ are found to remain correlated along the *c*-axis even at 400 K in **I**. We then present the measured magnetic susceptibility in the 230 to 290 K range to verify that the $NH_4^+$ only become ordered along the *c*-axis in **V**, i.e. there is no involvement of the $NH_4^+$ becoming ordered along the *c*-axis during the **IV-II** phase transition. Lastly, we discuss how the identification of $NO_3^-$–$NO_3^-$ magnetic interactions in AN allows for new opportunities to experimentally study the lattice dynamics of the vast array of solids possessing planar ions.

EXPERIMENTAL

For the synthesis of ammonium nitrate crystals, one equivalent 68% of nitric acid was slowly added to one equivalent 25-28% of ammonium hydroxide aqueous solution (pH level near 5). The resulting ammonium nitrate solution was stirred for 30 minutes to form ~4.3 moles of ammonium nitrate solution. The solution was then slowly left to evaporate for one week. The obtained crystals were then recrystallized at least three times in an $H_2O$-based solution with supersaturation rates of ≤0.03 at 300 K. The crystals were translucent, flat and elongated along the *a*-axis direction. Obtained single crystal data is available in the Supplementary Material. The masses of the measured samples varied from 1.3 to 14.8 mg. We only observed phase **III** around 10% of the time where clear discontinuities occurred between 305 K and 321 K in the magnetic susceptibility.

Hazard warning: while ammonium nitrate is stable under ambient conditions, it should not be stored near combustible materials as a strong enough initiation charge can lead to detonation. Only well-ventilated areas are appropriate when handling the solution and solid samples. Ammonium nitrate is not hazardous to health but proper protection equipment should still be worn to protect the eyes, face, hands and body.

The molar magnetic susceptibilities were derived from the measured magnetization of the single crystals with the Vibrating Sample Magnetometer (VSM) option of a Physical Properties Measurement System (PPMS) unit fabricated by Quantum Design, Inc. The samples were secured by GE Varnish onto quartz sample rods for the measurements.

RESULTS AND DISCUSSION

Figures 2(a), 2(b) and 2(c) show the magnetic susceptibilities $\chi_a(T)$, $\chi_b(T)$ and $\chi_c(T)$ with respect to temperature of AN under an applied magnetic field of $H = 1$ T for the cases when $H$ was aligned along the $a$-, $b$- and $c$-axis of the crystal lattice, respectively, in the 300 to 400 K region. The sweeping rate of the curves were 0.5 K/min.

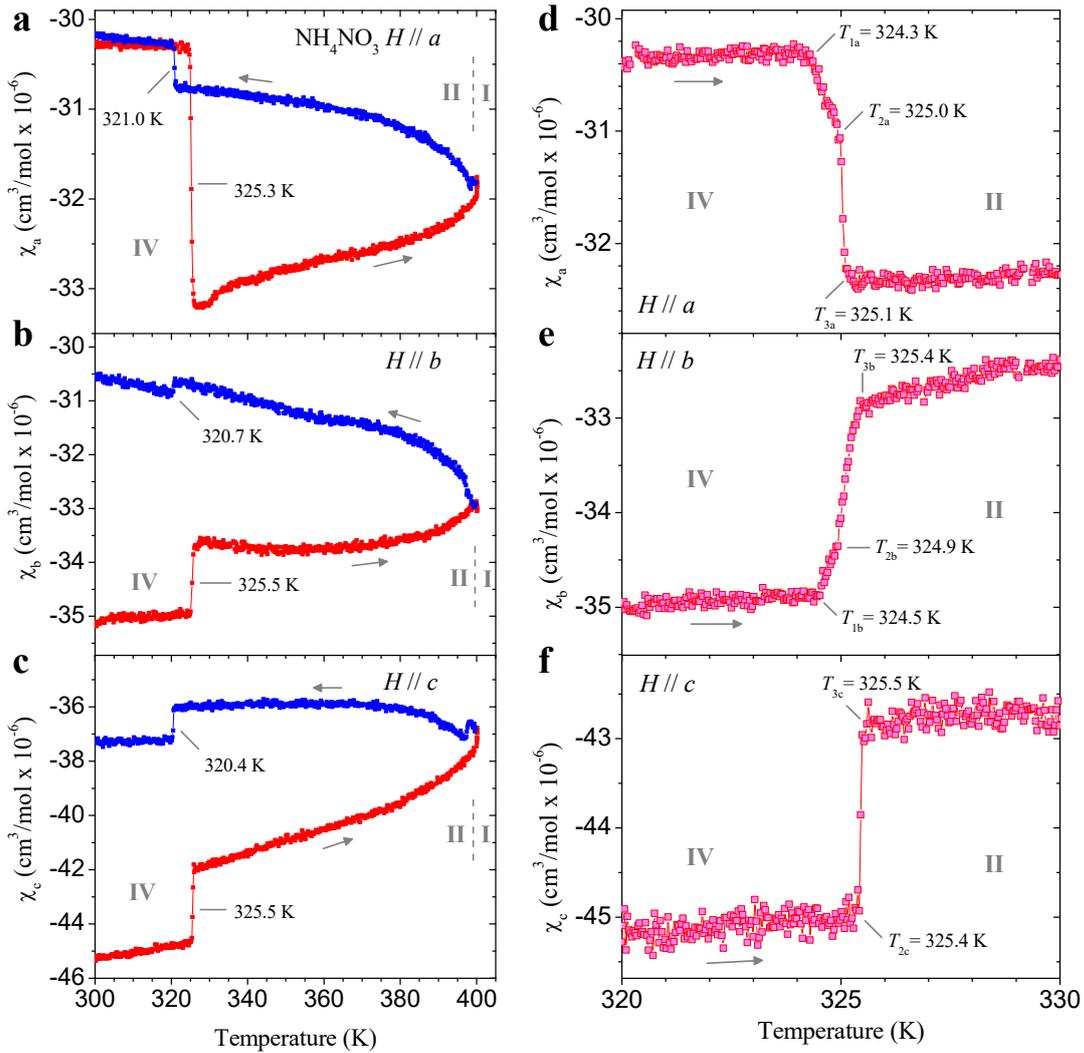

**FIG. 2.** Molar magnetic susceptibilities $\chi_a(T)$, $\chi_b(T)$ and $\chi_c(T)$ of NH$_4$NO$_3$ single

crystals along the (a) *a*-axis, (b) *b*-axis and (c) *c*-axis during warming and cooling in the 300–400 K range, respectively. Applied magnetic field was $H = 1$ T and sweeping rates were 0.5 K/min. (d) $\chi_a(T)$, (e) $\chi_b(T)$ and (f) $\chi_c(T)$ in the vicinity of $T_{IV-II}$ during warming under 0.1 K/min. Critical points $T_1$, $T_2$ and $T_3$ help distinguish the two segments comprising the sharp changes of $\chi_a(T)$ and $\chi_b(T)$ when phase transitioning from **IV** to **II**.

At $T_{IV-II} = 325.3$ K, $\chi_a(T)$ exhibited a sharp drop where the magnitude increased by 10.6%. At nearly the same temperature of 325.5 K, $\chi_b(T)$ and $\chi_c(T)$ also exhibited sharp changes but they were step-up discontinuities where the magnitude decreased by 3.8 and 6.1%, respectively. The observed transition temperatures are in good agreement with the order-disorder phase transition temperature from **IV** to **II** of previous reports employing other methods.[3,8,13,35-38]

When cooling back from **II** to **IV**, coinciding discontinuities occurred between 320.4 to 321.0 K. Along all three axis orientations, the changes in magnitude of $\chi(T)$ across $T_{II-IV}$ were less pronounced; we reckoned that this was because the *a*-axis of the crystal has a probability of remaining as the *a*-axis or transforming into the *b*-axis when undergoing the tetragonal-to-orthorhombic structural phase transition at $T_{II-IV}$.[3]

The phase transition between **II** and **I** in AN occurs near 398 K. Since the upper limit of our magnetometer is 400.0 K, pronounced changes in $\chi_a(T)$, $\chi_b(T)$ and $\chi_c(T)$ were not observed upon warming. However, after waiting at 400.0 K for the phase transition to finish, a small step was observed in $\chi_c(T)$ at $T_{I-II} = 397.5$ K upon cooling. A less pronounced step occurred at $T_{I-II}$ in $\chi_a(T)$ while $\chi_b(T)$ exhibited a small sudden increase. Discontinuities of such small magnitudes or no discontinuity at all are what is typically expected when a system undergoes a structural phase transition with no magnetic ordering. Consequently, the large changes in $\chi_a(T)$, $\chi_b(T)$ and $\chi_c(T)$ across $T_{IV-II}$, especially when the specific volume only increases by 1.69 to 2.00%,[3,11,13] is direct evidence of $NH_4^+$–$NH_4^+$ and $NO_3^-$–$NO_3^-$ magnetic disordering occurring at the **IV-II** phase transition.

A closer inspection of $\chi_a(T)$ across $T_{IV-II}$ by employing a warming speed of only 0.1 K/min revealed that the sharp drop is actually comprised of two steps with distinct slopes [Fig. 2(d)]. Under the same sweeping rate $\chi_b(T)$ was observed to also be comprised of two steps, but in the reverse direction [Fig. 2(e)]. In contrast, $\chi_c(T)$ only possessed one step anomaly [Fig. 2(f)] throughout the entire transition. The transition temperatures $T_1$, $T_2$ and $T_3$ are employed to label the points at which the changes of slopes occur to better gauge the duration of the two processes. At least five samples were measured for each of the three principal axes and the warming parts of the curves in Fig. 2 are highly reproducible provided the sample is always freshly grown and dry (*i.e.* did not phase transition to **III** in the temperature range of 305 < *T* < 318 K due to occluded $H_2O$) and starting from 300 K. The critical points varied by ±0.2 K across all samples.

The diamagnetic contributions of the $NH_4^+$ and $NO_3^-$ ions are $-13.3 \times 10^{-6}$ and $-18.9 \times 10^{-6}$ cm$^3$/mol,[39] respectively, so the expected diamagnetic susceptibility stemming from all of the paired electrons in AN is $\chi_e = -32.2 \times 10^{-6}$ cm$^3$/mol. In phase **II**, both $\chi_a$ and $\chi_b$ are nearly equal to $\chi_e$ confirming both the anions and cations are disordered along the *a* and *b*-axes. In contrast, $\chi_c(T)$ was slightly more negative than $\chi_e$ suggesting that the anions remained weakly correlated along the *c*-axis. The continuous uptick of $\chi(T)$ along all axes upon crossing **II** agrees with proton spin relaxation studies concluding that the $NH_4^+$ and $NO_3^-$ ions successively become more disordered with increasing temperature up to the melting point.[6] Even upon reaching phase **I** at 400.0 K, $\chi_c(T)$ did not reach $\chi_e$ indicating that the anions do not yet fully disorder at this high a temperature. This result appears reasonable since the $NO_3^-$ form columns along the *c*-axis.[40] Upon cooling across phase **II**, $\chi(T)$ along all three axes did not retrace their warming curves. This result may be due to phases **II** and **I** of ammonium nitrate being classified as a brittle and plastic crystal, respectively, according to its diffusion coefficient.[6] Samples usually crack upon crossing the **IV-II** phase transition so as the samples are cycled the cooling and warming curves in these two phases highly depend on the history of the sample.

In phase **IV**, $\chi_a(T)$ is smaller in magnitude than $\chi_e$ indicative of the cations being the predominant species exhibiting magnetic interactions along the *a*-axis. The $NO_3^-$ should also be ordered along the *a*-axis, however their associated signal appears to be masked by the larger signal stemming from the $NH_4^+$–$NH_4^+$ interactions. Hence, at $T_{IV-II}$, $\chi_a(T)$ should exhibit a step-down and a step-up component associated to the disordering processes of the $NH_4^+$ and $NO_3^-$, respectively (Figure 3). In addition, the faces of the $NO_3^-$ planes, which face the *b*-axis in **IV**, rotate by 45° about the *c*-axis when phase transitioning to **II** (as concluded in Ref. 3) so the direction perpendicular to the $NO_3^-$ planes shifts away from the *b*-axis by 45° and are projected onto the *a*-axis. Such shifting of the $NO_3^-$ planes is equivalent to a permanent magnetic moment reorientation of $\mu_{NO3}$ so $\chi_a(T)$ should possess an additional step-down component upon crossing $T_{IV-II}$. These individual contributions are summarized in Fig. 3 and allow us to hypothesize the reason why two steps were observed in $\chi_a(T)$ at $T_{IV-II}$. We posit that at the transition point $T_1$, the $NO_3^-$ planes begin to shift away from the *b*-axis only to finish at either $T_2$ or $T_3$. As this process takes place, both the $NH_4^+$ and $NO_3^-$ only begin to disorder at $T_2$, which explains the sequence of two drops observed in $\chi_a(T)$ with the latter having a larger negative slope.

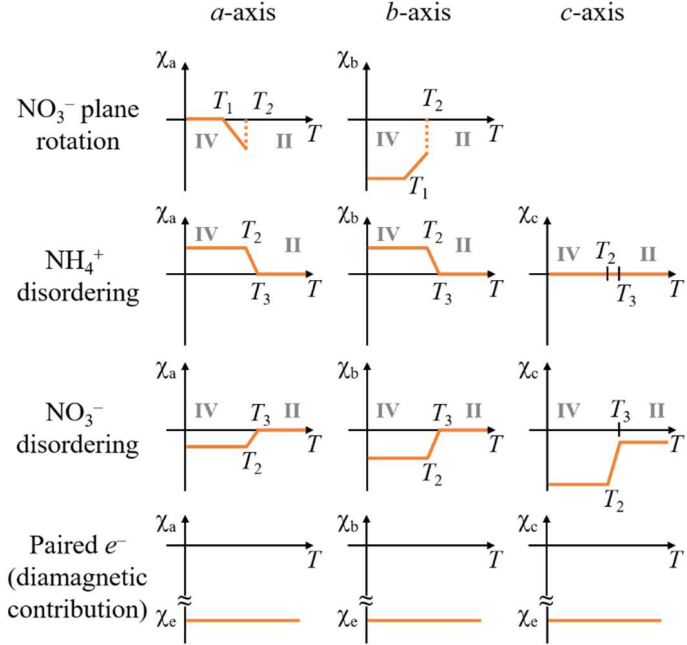

**FIG. 3.** Individual components contributing to $\chi(T)$ along the $a$-, $b$- and $c$-axes when crossing $T_{\text{IV-II}}$. The curves shown in Fig. 2 can be obtained by adding the corresponding columns of this figure.

By the same token, upon crossing $T_{\text{IV-II}}$, $\chi_b(T)$ should exhibit one step-down component arising from the disordering of the $NH_4^+$ and two step-up components, one stemming from the disordering of the $NO_3^-$ and another due to the shifting of the $NO_3^-$ planes away from the $b$-axis by 45°. Again, the shifting of the $NO_3^-$ planes occurs from $T_1$ to $T_2$ or $T_1$ to $T_3$ and the disordering process occurs from $T_2$ to $T_3$. Addition of these three components reproduces the two successive sharp increases observed in $\chi_b(T)$ with the latter one having a larger positive slope.

For the case of $\chi_c(T)$, only one sharp step was observed at $T_{\text{IV-II}}$ even at the slow warming rate of 0.1 K/min suggesting that the shifting of the $NO_3^-$ planes at $T_{\text{IV-II}}$ is strictly confined about the $c$-axis, *i.e.* the normal vector of the $NO_3^-$ planes, the direction of $\mu_{NO3}$, remain in the $ab$-plane during the process. Hence, the overall change in $\chi_c(T)$ at the phase transition should only be comprised of a step-down component and a step-up component arising from the disordering of the cations and anions, respectively. However, given the large magnitude of $\chi_c$ in both phases **IV** and **II**, it appears that the $NH_4^+$ are not ordered along the $c$-axis in these two phases and possibly only become ordered in **V**. To verify this, the magnetic susceptibility was also measured down to 230 K.

Figures 4(a), 4(b) and 4(c) show $\chi_a(T)$, $\chi_b(T)$ and $\chi_c(T)$ of AN first cooled from 300 K down to 230 K and then warmed back to 300 K at 0.5 K/min. At the **IV-V** phase transition, $\chi_a(T)$ exhibited a sharp increase in magnitude by 11.2%; in contrast, $\chi_b(T)$ and $\chi_c(T)$ decreased in magnitude by 5.8% and 7.7%, respectively. In the reverse direction going from **V** to **IV**, the magnitudes of $\chi_a(T)$ and $\chi_c(T)$ did not fully return to their original values which can be understood given the fact that the system does not fully convert to **IV** but rather becomes an admixture of **IV** and **V** (Refs. 13,16) upon returning to room temperature. The critical temperatures are labelled in Figures 4(a), 4(b) and 4(c) and they are in good agreement with those reported in existing literature.[3-5, 11,13,16] We also measured $\chi_a(T)$, $\chi_b(T)$ and $\chi_c(T)$ crossing the **IV-V** phase transition under the slow sweeping rate of 0.1 K/min [Figs. 4(d), 4(e) and 4(f)] to check if whether this transition was similar to the **IV-II** phase transition, however, along all directions only one abrupt discontinuity was observed.

At the **IV-V** phase transition, the specific volume only changes by 3.05 to 3.2% (Refs. 11,13) so the large changes in the magnetic susceptibilities clearly pertain to changes in the degree of magnetic ordering of the system. Similar to the **IV-II** phase transition, the principal axes do not change at the **IV-V** phase transition; the main changes are the doubling of the unit cell and the rotation of the $NO_3^-$ planes by 45°. From such, the step-down and step-up discontinuities in $\chi_a(T)$ and $\chi_b(T)$ when crossing **IV-V** are explainable as they should mirror the case when crossing the **IV-II** phase transition. Presuming that the rotation of the faces of the $NO_3^-$ planes also remain in the *ab*-plane, the step-up discontinuity in $\chi_c(T)$ can only be due to the ordering of the $NH_4^+$ along this axis. From this we can deduce that **IV** is partially disordered, in agreement to the conclusions of previous investigators,[3,17,18] however, we can pinpoint the $NH_4^+$ as not being ordered only along the *c*-axis in **IV**.

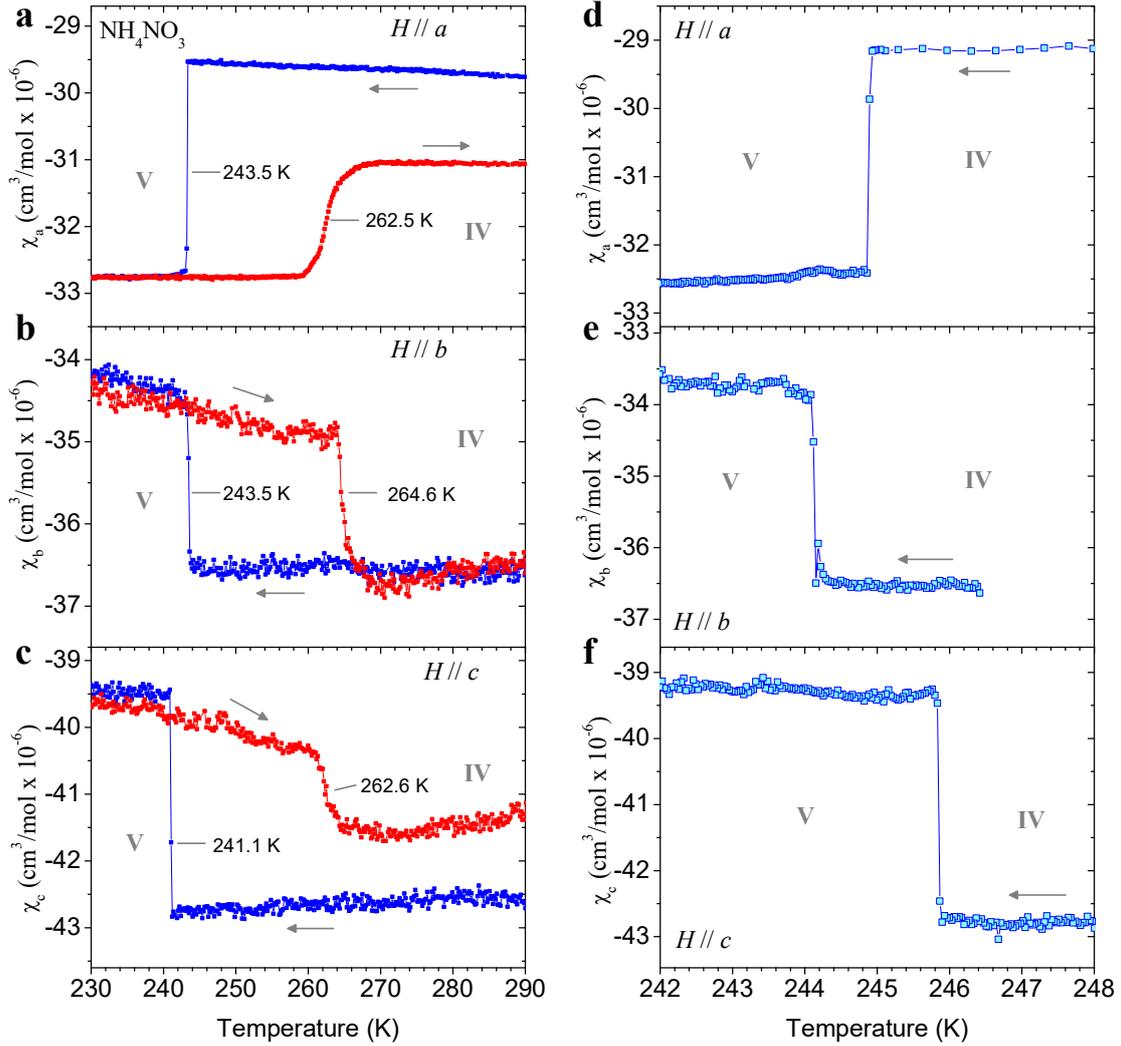

**FIG. 4.** $\chi_a(T)$, $\chi_b(T)$ and $\chi_c(T)$ of $NH_4NO_3$ single crystals along the (a) *a*-axis, (b) *b*-axis and (c) *c*-axis during warming and cooling below room temperature, respectively. Sweeping rates were 0.5 K/min and $H$ = 1 T. (d) $\chi_a(T)$, (e) $\chi_b(T)$ and (f) $\chi_c(T)$ in the vicinity of $T_{IV-V}$ during cooling under 0.1 K/min.

During a spin reorientation transition in ordinary magnetic systems a step-up and step-down discontinuity occurs in $\chi(T)$ when measured along different directions.[33,34] For instance, the $Fe^{3+}$ moments in $GdFe_3(BO_3)_4$ reorient from being antiferromagnetically ordered along the *a*-axis to the *c*-axis when cooled across its $T_{SR}$ = 9 K phase transition resulting in $\chi_a(T)$ and $\chi_c(T)$ to exhibit step-up and step-down discontinuities.[33] In the present case, the observation of two sharp steps in succession in $\chi_a(T)$ and $\chi_b(T)$ in $NH_4NO_3$ is evidence of a molecular analogue of the electron spin reorientation transition observed in conventional systems. We see that magnetic susceptibility measurements of highly dynamic systems comprised of planar ions can yield information on the extent of molecular gearing along the different crystallographic directions as well as the equilibrium orientations of the planar ions. This renders the currently employed method applicable to studying the lattice dynamics of the plethora of other nitrate salts, whether organic or inorganic for applications as phase change materials, as well as salts (and solid-state electrolytes) of other planar ions such as those of the carbonate, guanidinium, pyridinium, imidazolium and much more.[41-46]

CONCLUSION

In conclusion, the magnetic susceptibility of ammonium nitrate single crystals was carefully measured when crossing $T_{\text{IV-II}}$ (from phase **IV** to **II**) and $T_{\text{IV-V}}$ (from **IV** to **V**) along the three principal crystallographic orientations. Phase **IV** is not fully ordered in the sense that the $NH_4^+$ are not yet ordered along the *c*-axis until inhabiting **V**. When crossing $T_{\text{IV-II}}$, the rotation of the $NO_3^-$ planes about the *c*-axis by 45° occurs first before triggering the disordering of the $NH_4^+$ and $NO_3^-$ along the *a*- and *b*-axes. Upon inhabiting **II**, the system is not fully disordered since the $NO_3^-$ continue to remain weakly ordered along the *c*-axis columns. By "ordered" along the *c*-axis, it is meant that as an $NO_3^-$ rotates clockwise, its two immediate neighbors along the *c*-axis rotate counterclockwise by the same amount of degrees in a geared, hindered and random fashion[47] similar to how each $CH_3$ behaves in hexamethylbenzene at low temperatures[48-51] as well as amphidynamic crystals in their ordered phases.[52–56] According to the magnitudes of the magnetic susceptibilities $\chi_a$, $\chi_b$ and $\chi_c$, such correlated motion of the anions is only restricted to occur along the *c*-axis columns and does not extend along the *ab*-planes in **II**.

AUTHOR DECLARATIONS

The authors have no conflicts of interest to declare.

AUTHOR CONTRIBUTIONS

Du Na and Xintian Wang contributed equally to this work.

DATA AVAILABILITY

The data supporting the findings of this study are available from Fei Yen upon reasonable request.

SUPPLEMENTARY MATERIAL

Crystal data of ammonium nitrate $NH_4NO_3$; Observation of phase transitions from **IV**

to **III**; Additional magnetic susceptibility data under external magnetic field of 7 T.